\documentstyle[12pt,psfig,here]{article}
\newcommand{\be}{\begin{equation}}
\newcommand{\ee}{\end{equation}}
\newcommand{\bea}{\begin{eqnarray}}
\newcommand{\eea}{\end{eqnarray}}
\newcommand{\bean}{\begin{eqnarray*}}
\newcommand{\eean}{\end{eqnarray*}}
\newcommand{\ba}{\begin{array}}
\newcommand{\ea}{\end{array}}

\newcommand{\sslash}[1]{\not{\!#1}}

\newcommand{\norsl}{\normalsize\sl}

\begin{document}

\begin{titlepage}
\title{Revisiting  $W \gamma$ production at RHIC}

\author{
H. Kawamura\\
\norsl  Theory Group, DESY\\
\norsl  Platanenallee 6 D 15738 Zeuthen, GERMANY\\
\\
Y. Kiyo\\
\norsl  Dept. of Physics, Tohoku University\\
\norsl  Sendai 980-8578, JAPAN\\
\\
J. Kodaira and K. Morii\\
\norsl  Dept. of Physics, Hiroshima University\\
\norsl  Higashi-Hiroshima 739-8526, JAPAN\\}

\date{}
\maketitle

\vspace*{1.5cm}

\begin{abstract}
{\normalsize
\noindent
We discuss $W \gamma$ production in polarized $p \, p$
collisions at RHIC energy.
We point out that the RHIC collider has two advantages 
over other hadron colliders to measure the
characteristic feature of $W \gamma$ production:
(1) the RHIC energy is not so high and (2) the polarized beams are
available.
We calculate the tree level cross section for $W \gamma$ production
using a generic spin basis for $W$ and
discuss both the angular dependence and spin correlation.
}

\end{abstract}

\begin{picture}(5,2)(-330,-600)
\put(2.3,-95){DESY 01-098}
\put(2.3,-110){HUPD-0107}
\put(2.3,-125){TU-627}
\end{picture}

\thispagestyle{empty}
\end{titlepage}
\setcounter{page}{1}
\baselineskip 18pt 
\section{Introduction}

Radiative weak boson production in hadronic collisions
has been extensively discussed for testing the electroweak
structure of the standard model.
Especially, the $W \gamma$ production has been the subject of
much theoretical interests since this process contains the gauge
boson trilinear coupling and develops the so-called radiation
zero (RAZ)~\cite{BM}.
The phenomena RAZ is a typical example which is sensitive
to the structure of electroweak interaction.
 
After the discovery of RAZ, the tree level~\cite{HR}
and higher order~\cite{BHZO} analyses were done
some years ago.
Since the RAZ occurs only at the tree level of the fundamental process,
the convolution of the partonic cross section
with the parton distribution functions (PDFs)
and the higher order QCD corrections to the fundamental
process might give important
effects and it is possible that RAZ is completely smeared out
in the physical hadronic cross sections.
There have been many efforts to calculate higher order
corrections~\cite{SN,O,BHKSZ} in the standard model
and phenomenological analyses~\cite{BHO} to utilize this process
to test the standard model
and/or find a signal from the model beyond the standard model~\cite{DS}.

Due to the $V-A$ structure of the $W$ boson interaction,
only the initial quarks which have definite helicities can
participate in the process.
Therefore an experiment with the polarized beams will be more
efficient to study this process~\cite{WSCY,HMHJ}.

Many works so far assumed rather high energy collisions.
However, a realistic experiment with the polarized beams
became available as the RHIC Spin Project whose center of mass
energy is around $\sqrt{s} \sim 500\ {\rm GeV}$.
In this article, we reanalyze the radiative weak boson production
process at the RHIC energy.
We argue that the experiment in the RHIC energy region
will be very efficient to study this process if sufficient
luminocity is achieved.
From the experimental point of view, the detection of this process
might not be so easy.
We must be sure that the $W$ boson is produced.
The easiest signal would be the measurement of the
momentum and angular distribution of the decay product, {\it e.g.}
lepton, from $W$ boson.
This process, the weak boson production followed by its decay,
is, in principle, a very complicated one.
However, the narrow width approximation
for the weak boson makes the situation very simple.
We can discuss the production process and decay process separately.
In the decay process, it is known that the distributions 
of decay products strongly depend on the
direction of the spin of the produced $W$ boson.
So, it will be very interesting to know the cross section for the
polarized $W$ boson production.
In this article, we present the polarized $W$ boson production
in a \lq\lq generic\rq\rq\ spin basis according to ref.~\cite{MP1}.
Although the authors in ref.~\cite{HR} have already presented the
contributions to the cross section from various spin configurations,
the decomposition of the cross section in this article is new and
will be more useful from the realistic point of view.
One can expect that the \lq\lq beamline basis\rq\rq\ 
proposed in ref.~\cite{MP1} will be the most optimal spin basis here since
the speed of the produced $W$ would be small at the RHIC energy.
We are mainly interested in the $W \gamma$ production
process, however, we present the results also for the $Z \gamma$ production
process.

Although the analyses in this article are based on the tree level
calculations, we will give a discussion on the higher order QCD
corrections which might be unimportant in the RHIC energy regime.

This paper is organized as follows.
In Section 2, we give the tree level cross section for the 
polarized $W$ and $Z$ boson production in arbitrary spin basis.
In Section 3, we calculate the hadronic cross section by
convoluting the parton level cross section with the PDFs.
Section 4 is devoted to a discussion on the results obtained in Section 3.
Finally we give a summary in Section 5.  

\section{Polarized $W^{\pm}$ and $Z$ production}

We discuss, in this section, the polarized weak boson production
at the parton level.
We demonstrate that the produced weak boson is predominantly polarized
in the direction of the beam when its speed is not so large.
 
The tree level amplitude for the process,
\[    q_1 (q) + \bar{q_2} (\bar{q}) \to V (V) + \gamma (k)\ ,\]
where $V = W^{\pm}\ {\rm or}\ Z$, reads in the Feynman gauge,
\bea
    M^{W^{\pm}}_{\lambda\lambda '}
      &=&  - \, \delta_{i_1 i_2}\, e^2 \, g_L^W \, \bar{v} (\bar{q})
      \left[ Q_1 (\gamma_L)_{\mu}
            \frac{\sslash{q}-\sslash{k}}{u} \gamma_{\nu}
           + Q_2 \gamma_{\nu} \frac{\sslash{q}-\sslash{V}}{t}
                 (\gamma_L)_{\mu} \right. \nonumber\\
      &+& \,\left. \frac{Q_2 - Q_1}{s - M_W^2} (\gamma_L)^{\rho}
                      \left\{  (V - k)_{\rho} g_{\mu\nu}
             - (V + P)_{\nu} g_{\rho\mu}+  (k + P)_{\mu} g_{\rho\nu}
            \right\} \right] u (q)\nonumber\\
       & & \qquad\qquad\qquad\qquad\qquad\qquad  \times \ 
          \epsilon^{\mu *}_{\lambda} (V) \epsilon^{\nu *}_{\lambda '} (k)\ .
\label{amp1W}\\
    M^{Z}_{\lambda\lambda '}
      &=&  - \, \delta_{i_1 i_2}\, e^2 \,\sum_{\tau = L , R}
         g_{\tau}^Z \, \bar{v} (\bar{q})
      \left[ Q_1 (\gamma_{\tau})_{\mu}
            \frac{\sslash{q}-\sslash{k}}{u} \gamma_{\nu}
           + Q_2 \gamma_{\nu} \frac{\sslash{q}-\sslash{V}}{t}
                 (\gamma_{\tau})_{\mu} \right]  u (q)\nonumber\\ 
      & & \qquad\qquad\qquad\qquad\qquad\qquad \times \ 
           \epsilon^{\mu *}_{\lambda} (V) 
              \epsilon^{\nu *}_{\lambda '} (k) \ .\label{amp1Z}
\eea
Here, $\delta_{i_1 i_2}$ is the color indices for the incoming
quarks, $e (>0)$ is the electromagnetic coupling constant
and $Q_i$ is the electric charge of quark $q_i$.
For the $Z$ boson production, $Q_1 = Q_2$.
We define $P = q + \bar{q}$,\ 
$(\gamma_{L/R})_{\mu} = \gamma_{\mu} \frac{1 \mp \gamma_5}{2}$
and $M_W$ is the $W$ boson mass.
$\epsilon^{\mu *}_{\lambda} (V)$ ($\epsilon^{\nu *}_{\lambda '} (k)$)
is the $W$ or $Z$ (photon) polarization vector with spin index $\lambda$
($\lambda '$).
The weak boson-to-quark couplings $g_L^W$ , $g_{L/R}^Z$ are given by,
\[   g_L^W = \frac{V_{q_1 q_2}}{\sqrt{2} \sin \theta_W}\quad  ,\quad
   g_L^Z = \frac{T_3^i}{\sin\theta_W \cos\theta_W}
           - Q_i \tan\theta_W \quad , \quad
     g_R^Z = - Q_i \tan\theta_W \ ,\]
where $\theta_W$ is the Weinberg angle,
$V_{q_1 q_2}$ is the Cabibbo-Kobayashi-Maskawa matrix
and $T_3^i$ is the third component of weak isospin of $q_i$.
The parton level invariants $s , t , u$ are defined as usual,
\[ s = ( q + \bar{q})^2 \ , \  t = (q - V)^2 \ , \ u = (q - k)^2 \ .\]
Using the equation of motion for external particles,
both eqs.(\ref{amp1W}) and (\ref{amp1Z}) can be compactly expressed as,
\bean
    \lefteqn{M^V_{\lambda\lambda '}
      =  - \, \delta_{i_1 i_2}\, e^2 \, \sum_{\tau = L , R} g_{\tau}^V \, 
                \frac{Q_1 t + Q_2 u}{t + u}} \\
     & & \times \bar{v} (\bar{q})
    \left[ 2\, \left( \frac{q_{\mu}}{t} - \frac{\bar{q}_{\mu}}{u} \right)
            \gamma_{\nu} - \frac{2}{t} (V_{\nu} \gamma_{\mu} -
    g_{\mu\nu} \sslash{V})   
        - \frac{t+u}{tu} \sslash{V} \gamma_{\mu} \gamma_{\nu} \right]
     \omega_{\tau} u (q)
      \epsilon^{\mu *}_{\lambda} (V) \epsilon^{\nu *}_{\lambda '} (k)\ ,
\eean
with $\omega_{L/R} = \frac{1 \mp \gamma_5}{2}$.
The particular combination of charge and kinematical variables,
\[ \frac{Q_1 t + Q_2 u}{t + u} \ ,\]
leads to the phenomena called Radiation Zeros
for the $W$ production.

Assuming that we will measure only polarized $V$
boson, we sum up the photon polarization in the square of amplitude.
\be
  \frac{d \sigma^{\lambda}}{dt du}
   = \frac{1}{16 \pi s^2} \frac{1}{N_c^2} \sum_{{\rm color}, \lambda '}
        |M^V_{\lambda\lambda '}|^2 \ \delta (s + t + u - M_V^2 )\ .
\label{covx}
\ee
$N_c$ is the number of colors and $M_V$ denotes the mass of $W$ or $Z$.
In the zero momentum frame (ZMF), this expression becomes,  
\bea
  \frac{d \sigma^{\lambda}}{d \cos \theta}
      &=& \frac{1}{32 \pi  s} \left(1 - \frac{M_V^2}{s}\right)
      \frac{1}{N_c^2} \sum_{{\rm color}, \lambda '}
        |M^V_{\lambda\lambda '}|^2  \nonumber\\
   &\equiv& 
      \frac{e^4 \left( g_{L/R}^V \right)^2}{32 N_c \pi  s}
        \left(1 - \frac{M_V^2}{s}\right)
      \left(\frac{Q_1 t + Q_2 u}{t + u}\right)^2 
        \, \frac{1}{tu}\, {\cal S}^{\lambda}_{L/R}\ ,
\label{spinx}  
\eea
where the angle $\theta$ is the scattering angle
of the $V$ boson with respect to the quark $q_1$ direction.
\[ t = - \, \frac{s - M_V^2}{2} \, ( 1 - \cos \theta ) \quad , \quad
   u = - \, \frac{s - M_V^2}{2} \, ( 1 + \cos \theta ) \ .\]

In this paper, we use the spinor helicity method
with the generic spin basis proposed in ref.~\cite{MP1}.
According to ref.~\cite{MP2}, we introduce the spin vector $S$
such that $S^2 = -1$ and $V \cdot S = 0$.
The $V$ boson has spin projection $\epsilon^{\mu *}_{\lambda} (V)\ 
(\lambda = + , 0 , -)$ with respect to the spatial part of $S$
in the rest frame of $V$.
These three polarization vectors can be written in terms of
massless spinors with momenta,
$V_1 = ( V + M_V S )/2 \ , \ V_2 = ( V - M_V S )/2$ as,
\bean
     \epsilon^{\mu *}_{\pm} (V) &=& \frac{< V_1 \pm | \gamma^{\mu}
       | V_2 \pm >}{\sqrt{2} M_V} \ ,\\
     \epsilon^{\mu *}_{0} (V) &=& 
        \frac{< V_1 + | \gamma^{\mu}| V_1 + >
                  - < V_2 + | \gamma^{\mu}| V_2 + >}{2 M_V}
           = \frac{V_1^{\mu} - V_2^{\mu}}{M_V} \ .
\eean
The spin vector for $V$ is parameterized by
the angle $\xi$ in its rest frame measured from the $\gamma$ 
direction as given in Fig.1.
Choosing the z-axis to be the direction of $V$ momentum in the ZMF,
the momenta of particles and the vector $V_i$ turn out to be,
\bea
    q &=& \frac{\sqrt{s}}{2}\,( 1\,,\, \sin\theta \,,\,0 \,,\,
    \cos\theta ) \quad , \quad \bar{q} = \frac{\sqrt{s}}{2}\,
    ( 1\,,\, - \sin\theta \,,\,0 \,,\, - \cos\theta ) \ ,\nonumber\\
    V &=& M_V \gamma \,(1\,,\, 0 \,,\,0 \,,\, \beta ) \quad\quad , \ \quad\quad 
    k = M_V \gamma \beta \,(1\,,\, 0 \,,\,0 \,,\, 1 )\ ,\nonumber\\
    V_1 &=& \frac{M_V}{2}\,( \gamma (1 - \beta \cos \xi )\,,\,- \sin
    \xi\,,\, 0 \,,\, \gamma ( \beta - \cos \xi )) \ ,\label{vectors}\\
    V_2 &=& \frac{M_V}{2}\,( \gamma (1 + \beta \cos \xi )\,,\, \quad  \sin
    \xi\,,\, 0 \,,\, \gamma ( \beta + \cos \xi )) \ ,\nonumber
\eea
where $\beta$ is the speed of $V$ boson,
\[   \beta = \frac{s - M_V^2}{s + M_V^2} \ ,\]
and $\gamma = ( 1 - \beta^2 )^{-1/2}$ is the usual
relativistic factor.
\begin{figure}[H]
\begin{center}
\leavevmode\psfig{file=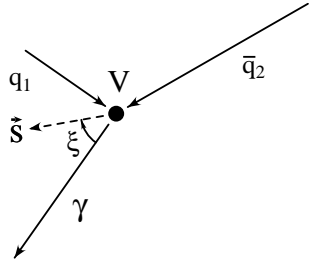,width=4cm}
\caption{$\vec{s}$ is the spin vector associated with $V$ in its
rest frame.}
\end{center}
\end{figure}

After straightforward manipulations, we obtain,
\bea
   {\cal S}^{0}_{L/R} &=& 2\, \left[ \, t^2 + u^2 + 2 M_V^2 s - 4 ( a^2
   + b^2 ) \, \right] \ ,
\label{spol}\\
   {\cal S}^{\pm}_{L/R} &=&  t^2 + u^2 + 2 M_V^2 s + 4 ( a^2
   + b^2 ) \, ( {\pm}/{\mp} ) \, 4 \left\{ (s + u ) \, a - ( s + t )\, b
   \right\} \ ,
\label{pmpol}
\eea
where we have defined,
\[  a = q \cdot (V_1 - V_2 )\ , \quad b = \bar{q} \cdot (V_1 - V_2 ) \ .\]
Summing over the three possible spins of $V$, we recover the usual
expression from eqs.(\ref{spol},\ref{pmpol}),
\[  \sum_{\lambda} {\cal S}^{\lambda}_{L/R}
     = 4\, \left[ \, t^2 + u^2 + 2 M_V^2 s \, \right] \ ,\]
for the unpolarized cross section which
is independent of the spin axis angle $\xi$, as it must be.
 
In the case of the hadronic reaction, we must finally convolute
the partonic cross section eq.(\ref{covx}) with the PDFs.
In such a case, the following two spin bases are sensible. 
Let us consider the $W^{\pm}$ production for definiteness.
In this case, only the left-hand coupling is present ($\tau = L$).
The first is the usual helicity basis ($\xi = \pi$).
Inserting $\xi = \pi$ to eq.(\ref{vectors}),
eqs.(\ref{spol},\ref{pmpol}) become,
\be
    {\cal S}^{0}_L = 4 M_W^2 s \, \sin^2 \theta \quad ,\quad
    {\cal S}^{\pm}_L = 2 M_W^2 s \,
            \frac{1 + \beta^2}{1 - \beta^2} (1 \mp \cos \theta )^2 \ .
\label{help}
\ee
From these expressions, one can see that all spin configurations
equally contribute to the cross section for finite $\beta$.  

The second is the so-called beamline basis~\cite{MP1},
which is defined by,
\be
   \cos \xi = \frac{\cos \theta + \beta}{1 + \beta \cos \theta}
    \quad , \quad \sin \xi = \frac{\sqrt{1 - \beta^2} \sin \theta}
    {1 + \beta \cos \theta} \ .
\label{beamline}
\ee
In this basis, the spin axis for $W$ is the direction of one of the
beams $\bar{q_2}$.
Inserting again eq.(\ref{beamline}) to eq.(\ref{vectors}),
we obtain in this case from eqs.(\ref{spol},\ref{pmpol}),
\bea
  {\cal S}^{0}_L &=& 8 M_W^2 s\,
          \frac{1}{( 1 + \beta \cos \theta )^2}\, \beta^2 \, \sin^2 \theta  \ ,
         \nonumber\\
  {\cal S}^{-}_L &=& 4 M_W^2 s\,
          \frac{1}{( 1 + \beta \cos \theta )^2}\, 
          \frac{\beta^4}{1 - \beta^2} \, \sin^4 \theta \ ,\label{beamp}\\
  {\cal S}^{+}_L &=& 4 M_W^2 s\,
          \frac{1}{( 1 + \beta \cos \theta )^2}\, 
          \frac{1}{1 - \beta^2} \, 
           \left\{ (1 - \beta^2 )^2 + ( 1 + \beta \cos \theta )^4
          \right\} \nonumber \ .
\eea
From these results, we can see that the the cross sections with the
polarization, $0$ and $-$ are suppressed when the speed $\beta$
of $W$ is small.
This is consistent with the following observation.
Due to the $V-A$ structure of the week interaction,
the spins of the initial quarks are aligned to the direction of  
$\bar{q_2}$ momentum.
Therefore near the threshold, only the $+$ component survives.

We present in Fig.2 (Fig.3) the differential cross section for $W \gamma$
($Z \gamma$) production at partonic level in the beamline
and helicity bases.
For the $Z$ production, we plot only the $q_L \bar{q}_R$ case.  
We take the center-of-mass energy to be $\sqrt{s} = 100\ {\rm GeV}$
which would be a typical energy scale of the partonic sub-process
in the RHIC energy regime.
\begin{figure}[H]
\begin{center}
\begin{tabular}{cc}
\qquad Beamline & \qquad Helicity \\
\leavevmode\psfig{file=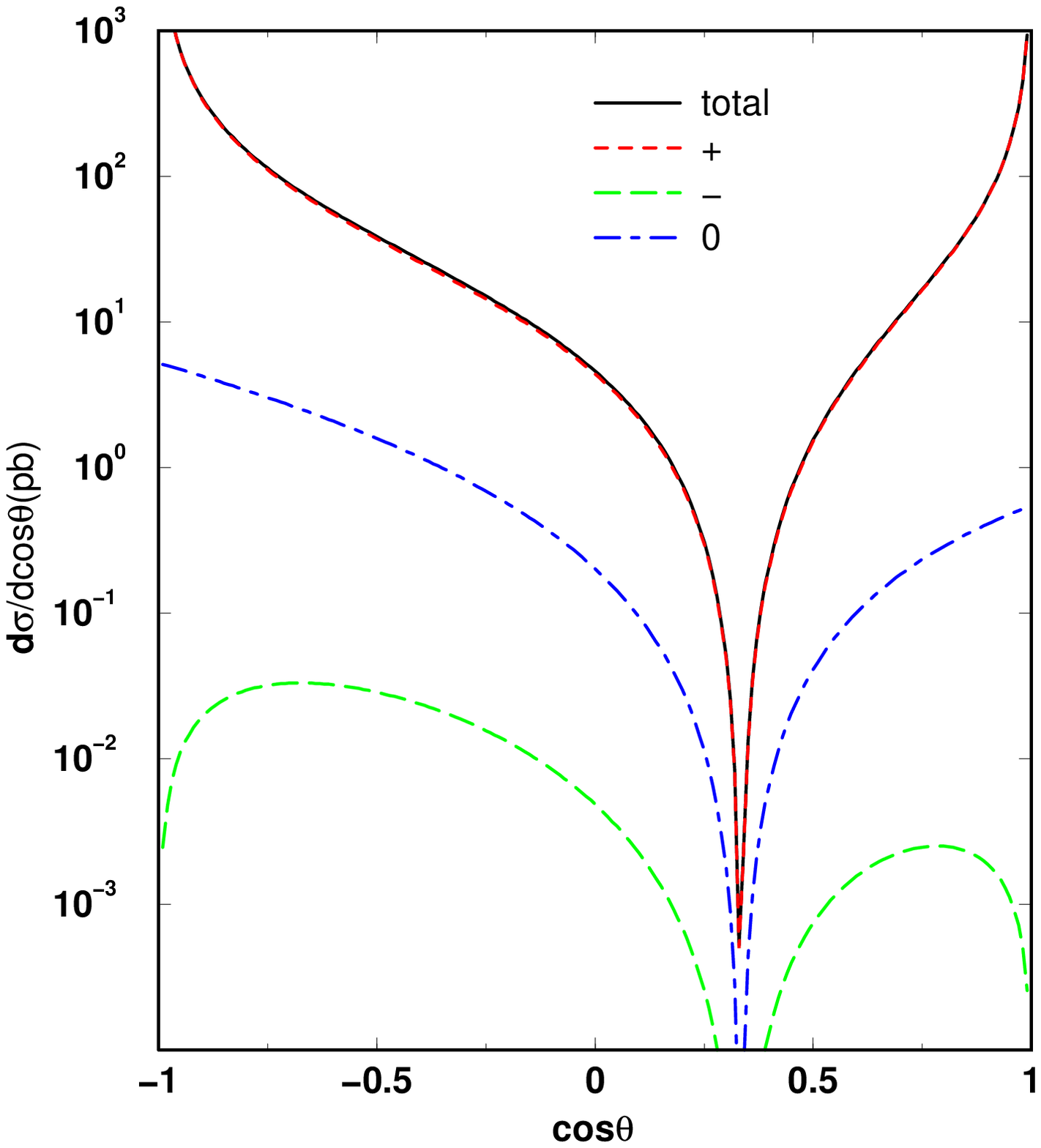,width=5cm,angle=-90}  &
\leavevmode\psfig{file=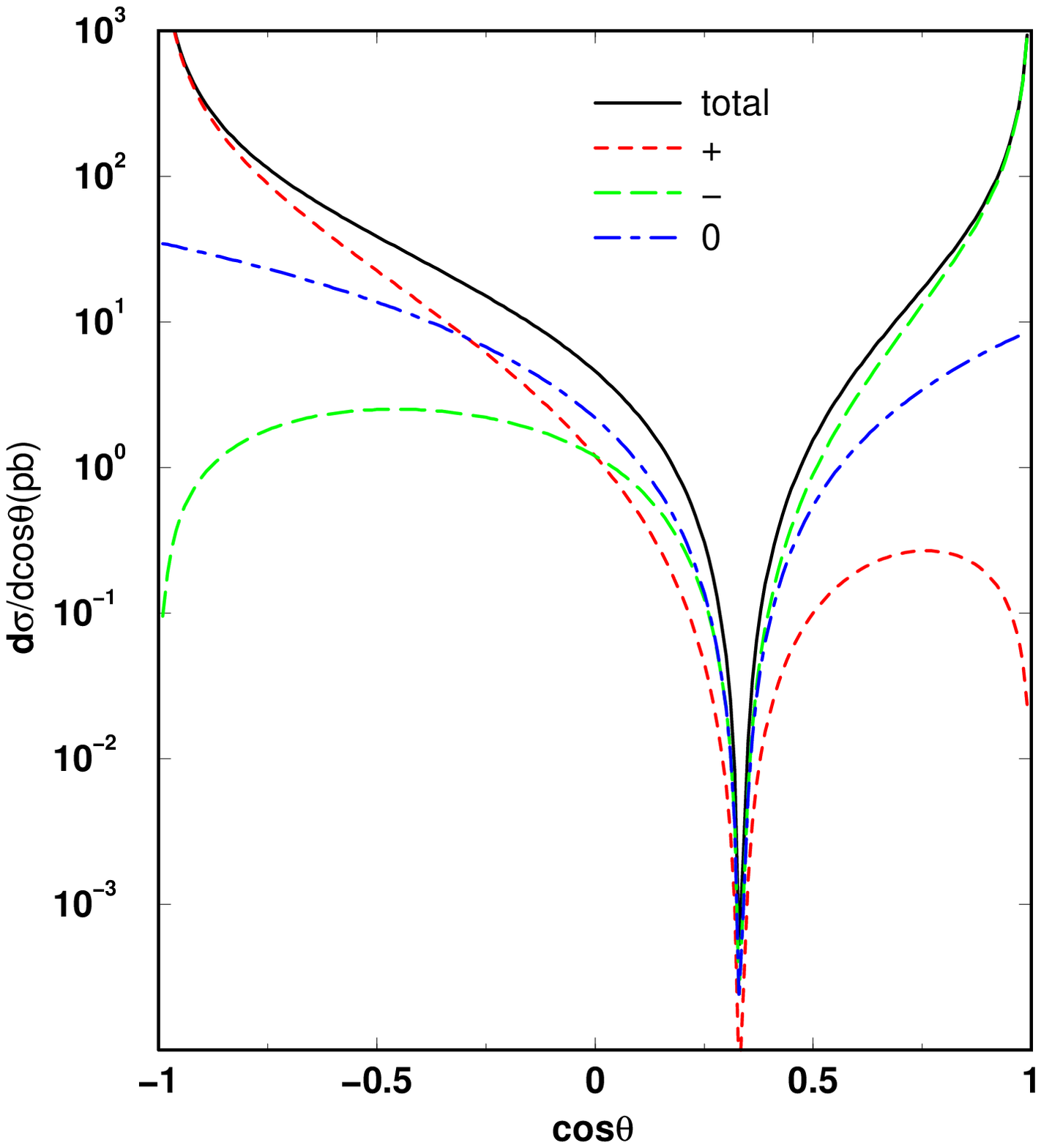,width=5cm,angle=-90}
\end{tabular}
\caption{$W \gamma$ production in the partonic ZMF.} 
\label{fig2} 
\end{center}
\end{figure}
\vspace{-0.8cm}
\begin{figure}[H]
\begin{center}
\begin{tabular}{cc}
\qquad Beamline & \qquad Helicity \\
\leavevmode\psfig{file=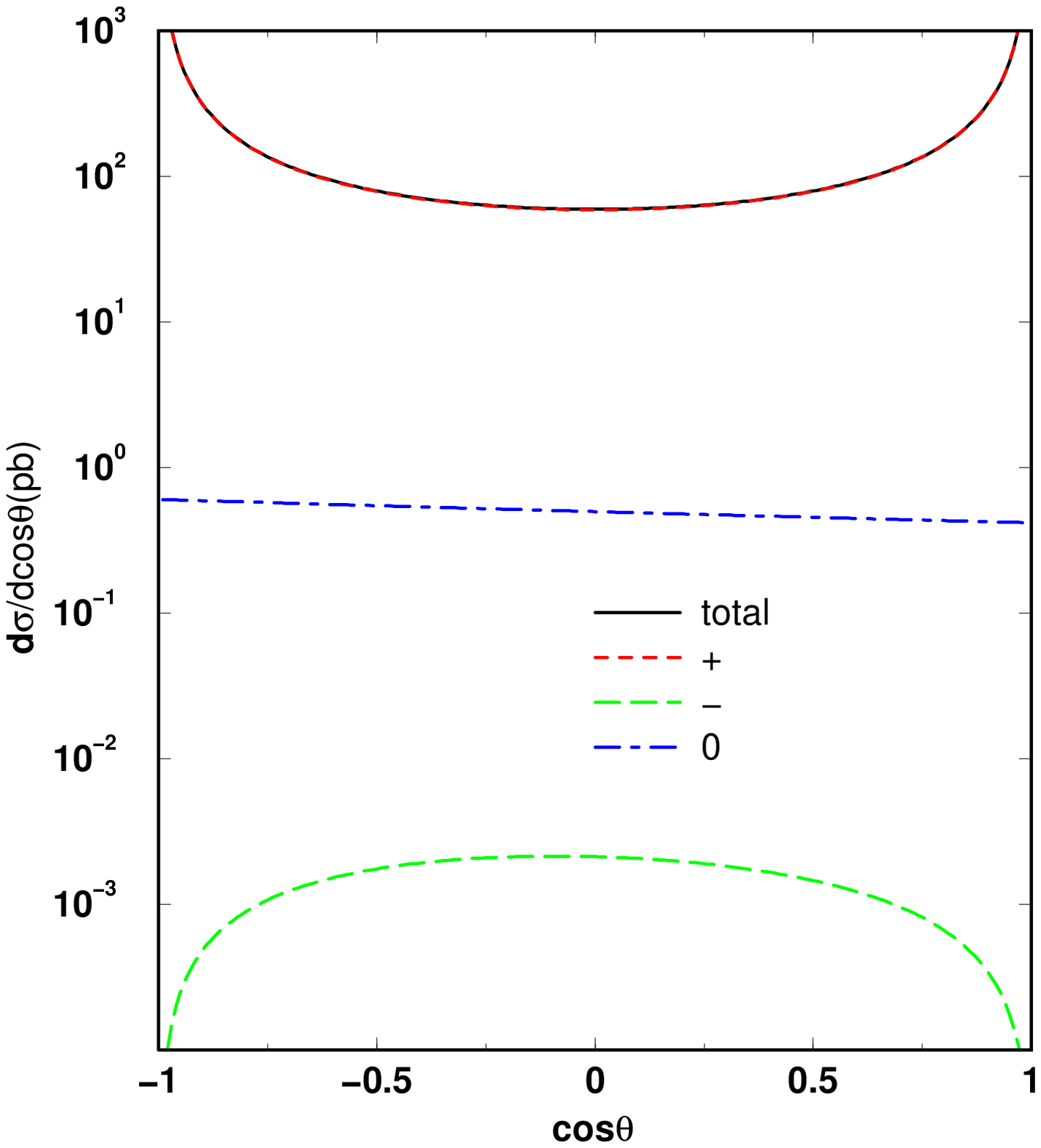,width=5cm,angle=-90}  &
\leavevmode\psfig{file=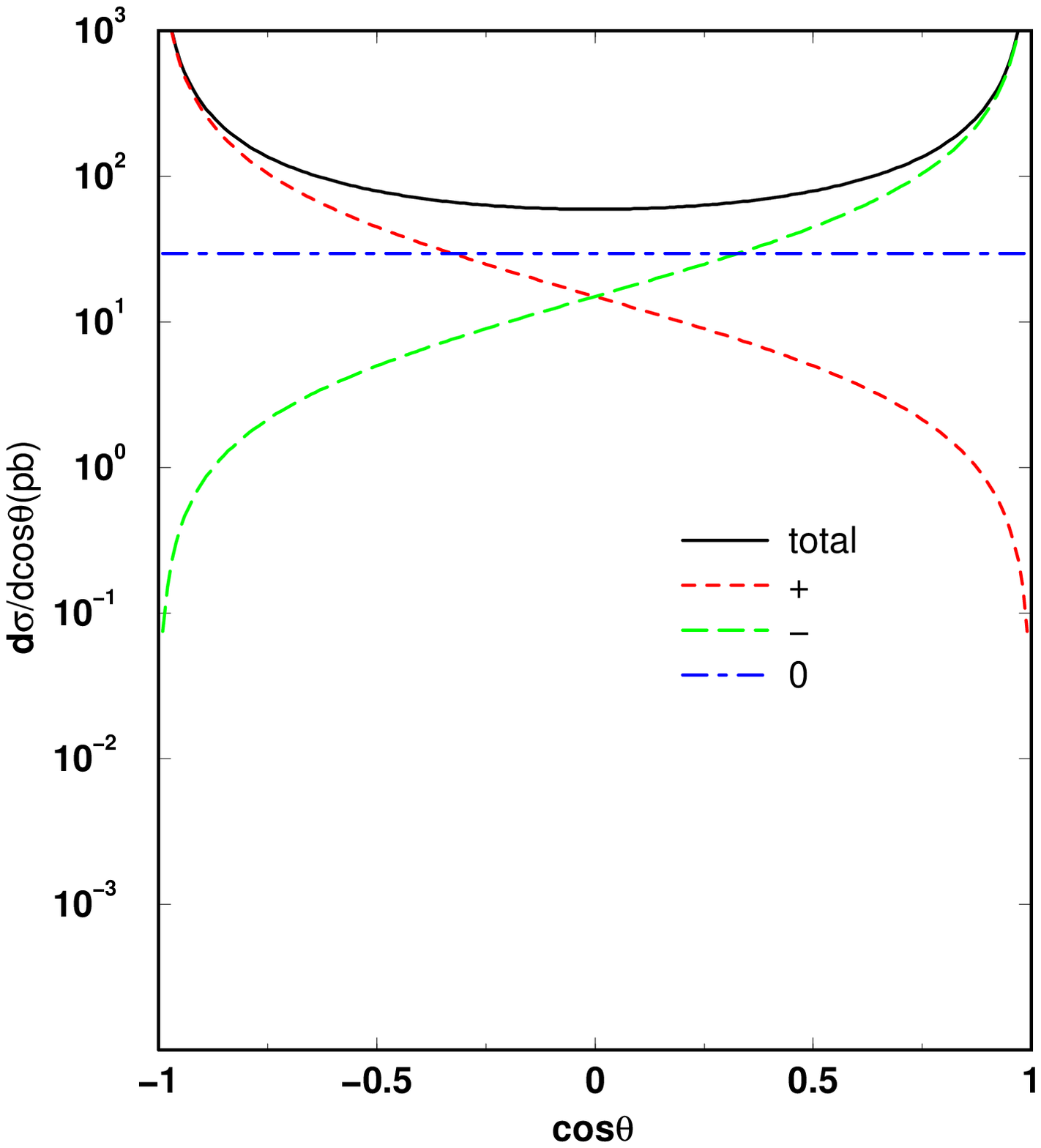,width=5cm,angle=-90}
\end{tabular}
\caption{$Z \gamma$ production in the partonic ZMF.} 
\label{fig3} 
\end{center}
\end{figure}
\noindent
Except for the fermion masses, which we neglect,
all input masses and coupling constants used in the numerical
computations are the central values as reported in the 2000
Review of Particle Physics~\cite{PDG}.

For both $W$ and $Z$ productions, only one component, namely $+$
component, dominates the cross section and contributions from
other spin configurations are less than several percents
in the beamline basis as expected since the $\beta$ is small,
$\beta \sim 0.2$.
This means that the direction of the spin of produced weak boson at
RHIC is almost the beam direction.
On the other hand, the helicity basis is a poor basis in the sense
that all components of spin contribute equally
to the cross section as seen in Figs.2. and 3.
For the $Z$ production, the helicity state of the quarks,
$q_R \bar{q}_L$ also contributes.
In this case, the polarization indices $\pm$ in
eqs.(\ref{help},\ref{beamp}) should be exchanged and the interpretation of
the results is the same as for the $W$ production case.

\section{Hadronic cross section}

We present the hadronic cross section
by convoluting the previous parton cross sections with the PDFs.
In this section, we shall limit ourselves to
the $W^+ \gamma$ production
and calculate the angular distribution of the produced
$\gamma$ as shown in Fig.4 simply due to
\begin{figure}[H]
\begin{center}
\leavevmode\psfig{file=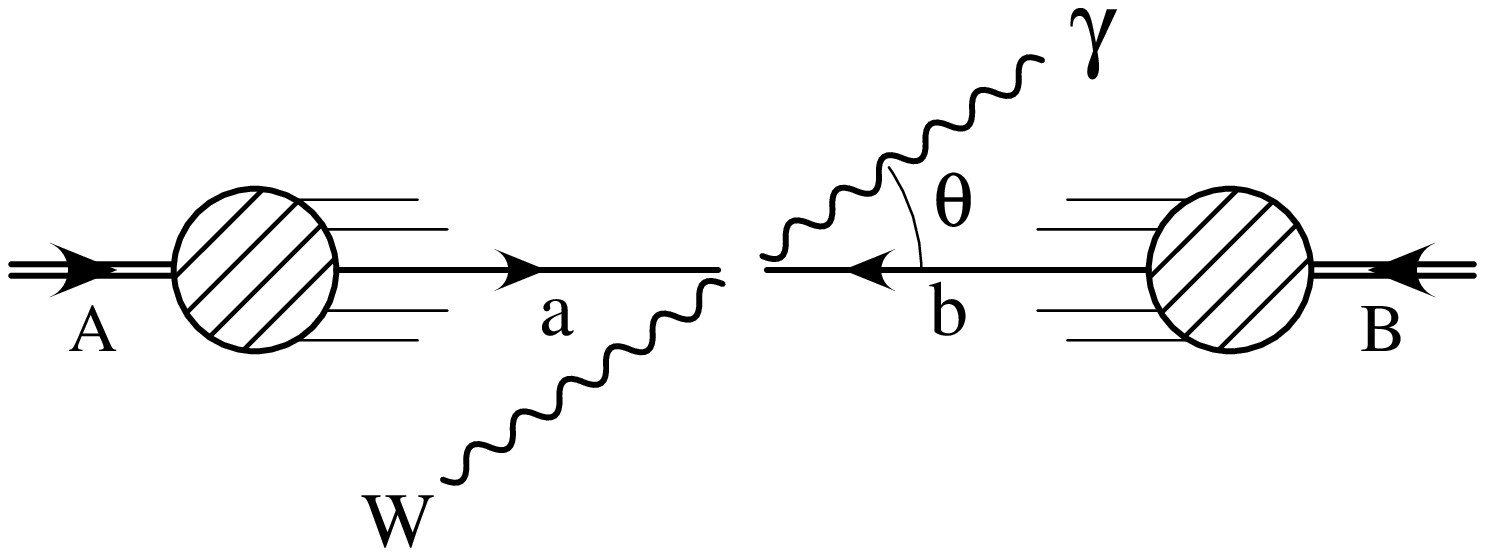,width=7cm}
\caption{The process $p p \to W \gamma X$.}
\end{center}
\end{figure}
\vspace{-0.3cm}
\noindent
the easier kinematics.
Furthermore we consider the spin-summed cross section
because we already know that the spin of the produced $W$
is in the beam direction. 

At first, we summarize the polarized quark distribution function
in the polarized proton and give qualitative arguments on
the results expected. 

\subsection{Parton distribution functions}

We consider only the contributions from $u$ and $d$ quarks
since the Cabibbo angle is $\sin^2 \theta_C \sim 0.05$
and the number of $s$ and $c$ quarks is expected to be small.
We use the notation $q_{\pm} (x,Q^2)$ for the quark distribution
in a polarized proton
with the momentum fraction $x$ at the scale $Q^2$,
where $+ (-)$ refers to a quark
with helicity parallel (antiparallel) to the parent proton's
helicity.
These quark densities roughly satisfy,  
\be
    u_+ (x,Q^2) \gg d_- (x,Q^2) \sim u_- (x,Q^2) \gg d_+ (x,Q^2)
        \gg \bar{u}_{+ , -} (x,Q^2)
         \sim  \bar{d}_{+ , -} (x,Q^2)\ , \label{distribution}
\ee
for almost all $x$ at the relevant scale $Q^2 \simeq M_W^2$.

The V-A structure of interaction picks up only quarks which
have definite helicities.
Therefore the partonic contents which contribute to
the hadronic cross section $d \sigma_{h_A h_B}$ where
$h_A,h_B = \pm$ denote the proton's helicities,
are the followings.
\bea
d \sigma_{++} &:& \quad
   u_- (x_A , Q^2) \bar{d}_+ (x_B , Q^2 ) + u \leftrightarrow \bar{d} \ ,
              \nonumber\\
d \sigma_{+-} &:& \quad
  u_- (x_A , Q^2) \bar{d}_- (x_B , Q^2 ) + 
      \bar{d}_+ (x_A , Q^2) u_+ (x_B , Q^2 ) \ ,
              \nonumber\\
d \sigma_{-+} &:& \quad
   u_+ (x_A , Q^2) \bar{d}_+ (x_B , Q^2 ) + 
      \bar{d}_- (x_A , Q^2) u_- (x_B , Q^2 ) \ , \label{combi}\\
d \sigma_{--} &:& \quad
  u_+ (x_A , Q^2) \bar{d}_- (x_B , Q^2 ) + u \leftrightarrow \bar{d} 
       \nonumber \ .
\eea
From these expressions and eq.(\ref{distribution}), we expect that:
(1) the size of the cross section is
$d \sigma_{--} > d \sigma_{+-} = d \sigma_{-+} > d \sigma_{++}$,
(2) the angular distributions become asymmetric for
$d \sigma_{+-}$ and $d \sigma_{-+}$ due to the RAZ of their
sub-processes.

\subsection{Covolution}

For the reaction,
\[  p (P_A) + p (P_B) \to W^+ (q) + \gamma (k) + X \ ,\]
the differential cross section for the photon can be derived from
the expression,
\be
  s\, \frac{d \sigma_{h_A h_B}}{dt du} = 
    \int dx_A dx_B f_{h_A}^i (x_A , Q^2) f_{h_B}^j (x_B , Q^2)
       \, \hat{s} \, 
      \frac{d \hat{\sigma}_{ij}}{d\hat{t} d\hat{u}}\ ,\label{hadronx}
\ee
where $f_{h_A , h_B}^i$ stands for one of 
the quark distributions, eq.(\ref{distribution}).
Here $d \hat{\sigma}_{ij}$ denote the partonic cross sections eq.(\ref{covx})
where $i (j)$ corresponds to a parton from proton A (B).
The Mandelstam variable in the hadron system are defined by,
\[  s = (P_A + P_B)^2\ ,\ t = (P_A - k)^2 \ , \ u = (P_B - k)^2 \ .\]
Therefore the partonic variables are,
\[  \hat{s} = x_A x_B s \ ,\quad \hat{t} = x_A t \ , \quad
                \hat{u} = x_B u \ .\]
From eqs.(\ref{hadronx}) and (\ref{covx}), the angular distribution of photon
in proton's ZMF becomes,
\bea
  \frac{d \sigma_{h_A h_B}}{d \cos \theta_{\gamma}} &=& 
    \frac{1}{8 \pi s}\int k^0 dk^0  \frac{dx_A}{x_A} \frac{dx_B}{x_B}
     f_{h_A}^i (x_A , Q^2) f_{h_B}^j (x_B , Q^2) \\ \nonumber
  & &  \qquad\qquad \times \frac{1}{N_c^2} \sum_{{\rm color}, \lambda\lambda '}
        |M^W_{\lambda\lambda '}|^2 \ 
     \delta (\hat{s} + \hat{t} + \hat{u} - M_W^2 )\ ,\label{photonx}
\eea
where $k^0$ is the energy of photon and it is understood that
$i,j$ take appropriate partons according to eq.(\ref{combi}).
 
We plot the hadronic cross section in Fig.5 at $\sqrt{s} = 500 {\rm GeV}$
using the parameterization in ref.~\cite{MRS} with $Q^2 = M_W^2$
for the (quark) parton densities.
To avoid the soft photon singularity or
take into account the experimental situation,
we have chosen the 
\begin{figure}[H]
\begin{center}
\begin{tabular}{cc}
\qquad $d \sigma_{+-}/d \cos \theta_{\gamma}$ & 
\qquad $d \sigma_{-+}/d \cos \theta_{\gamma}$ \\
\leavevmode\psfig{file=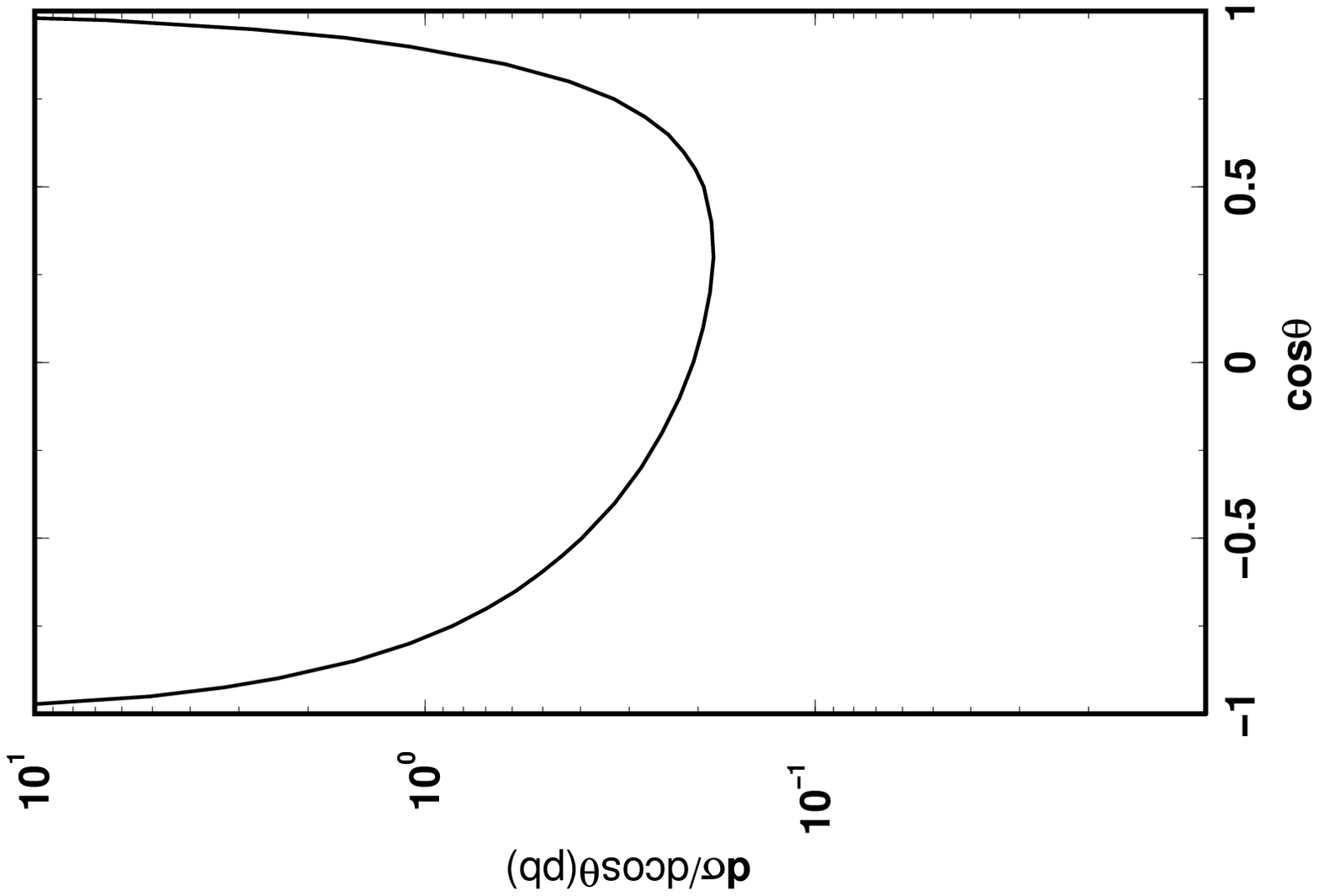,width=4cm,angle=-90} &
\leavevmode\psfig{file=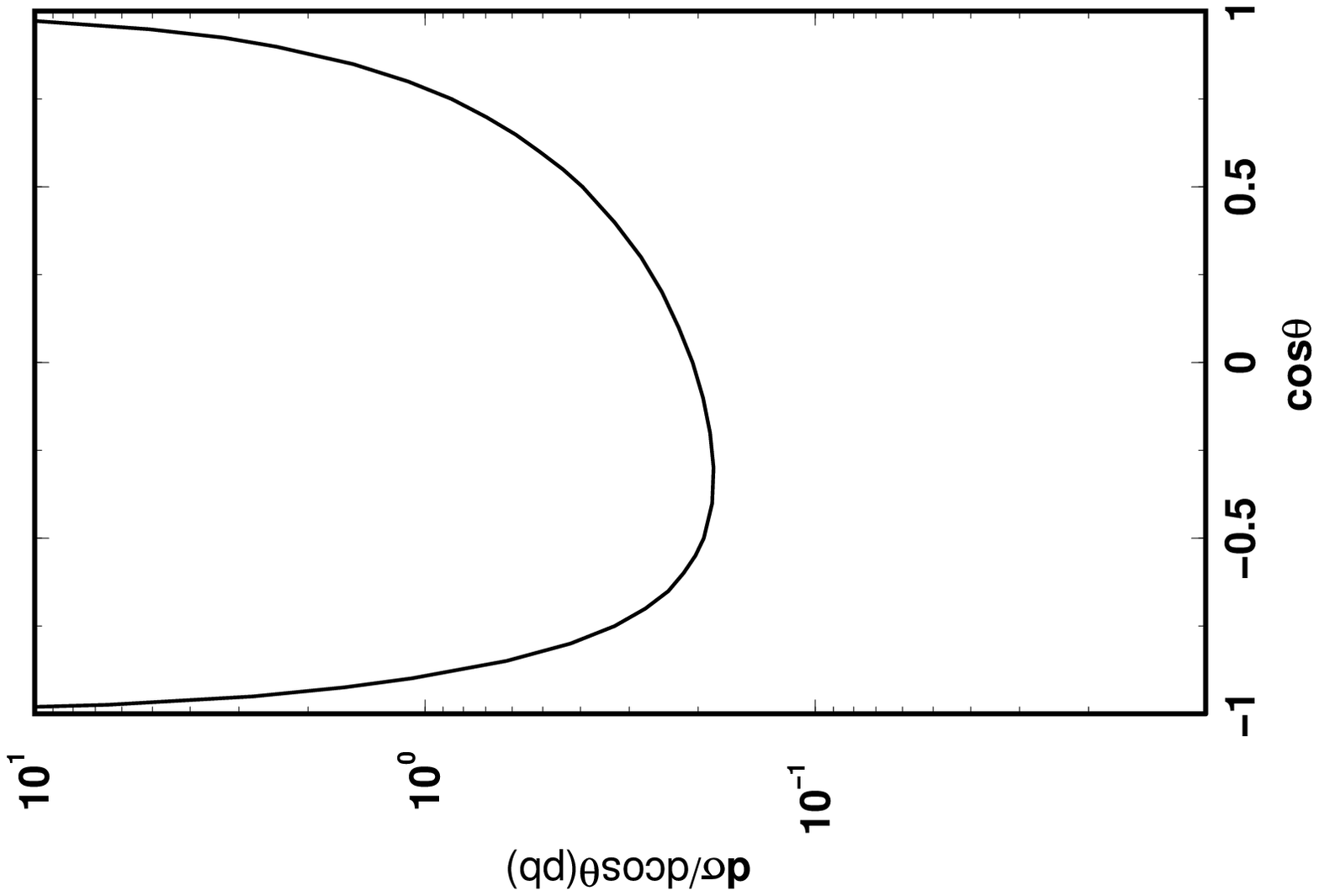,width=4cm,angle=-90}
\end{tabular}
\begin{tabular}{cc}
\qquad $d \sigma_{--}/d \cos \theta_{\gamma}$ &
\qquad $d \sigma_{++}/d \cos \theta_{\gamma}$ \\
\leavevmode\psfig{file=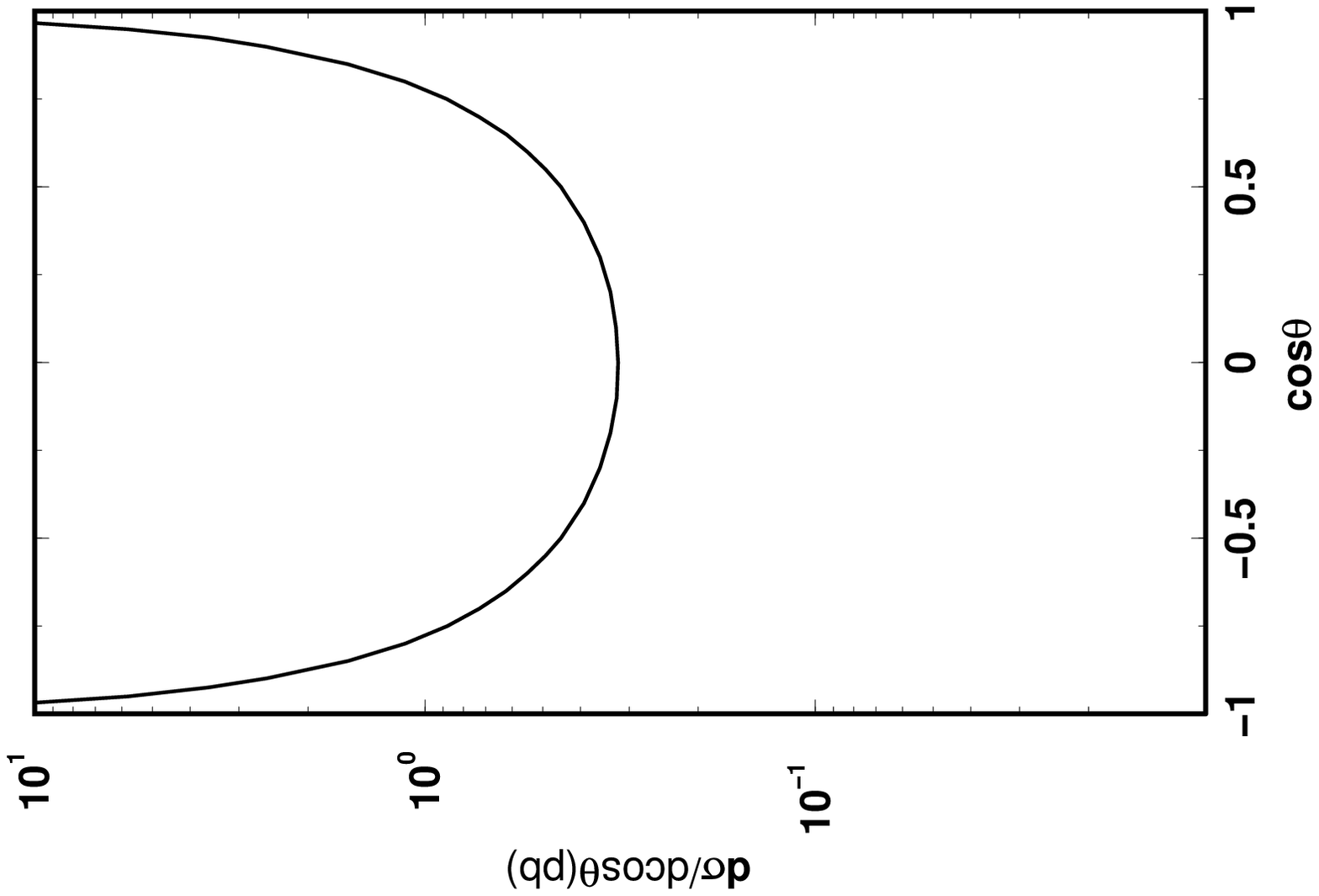,width=4cm,angle=-90} &
\leavevmode\psfig{file=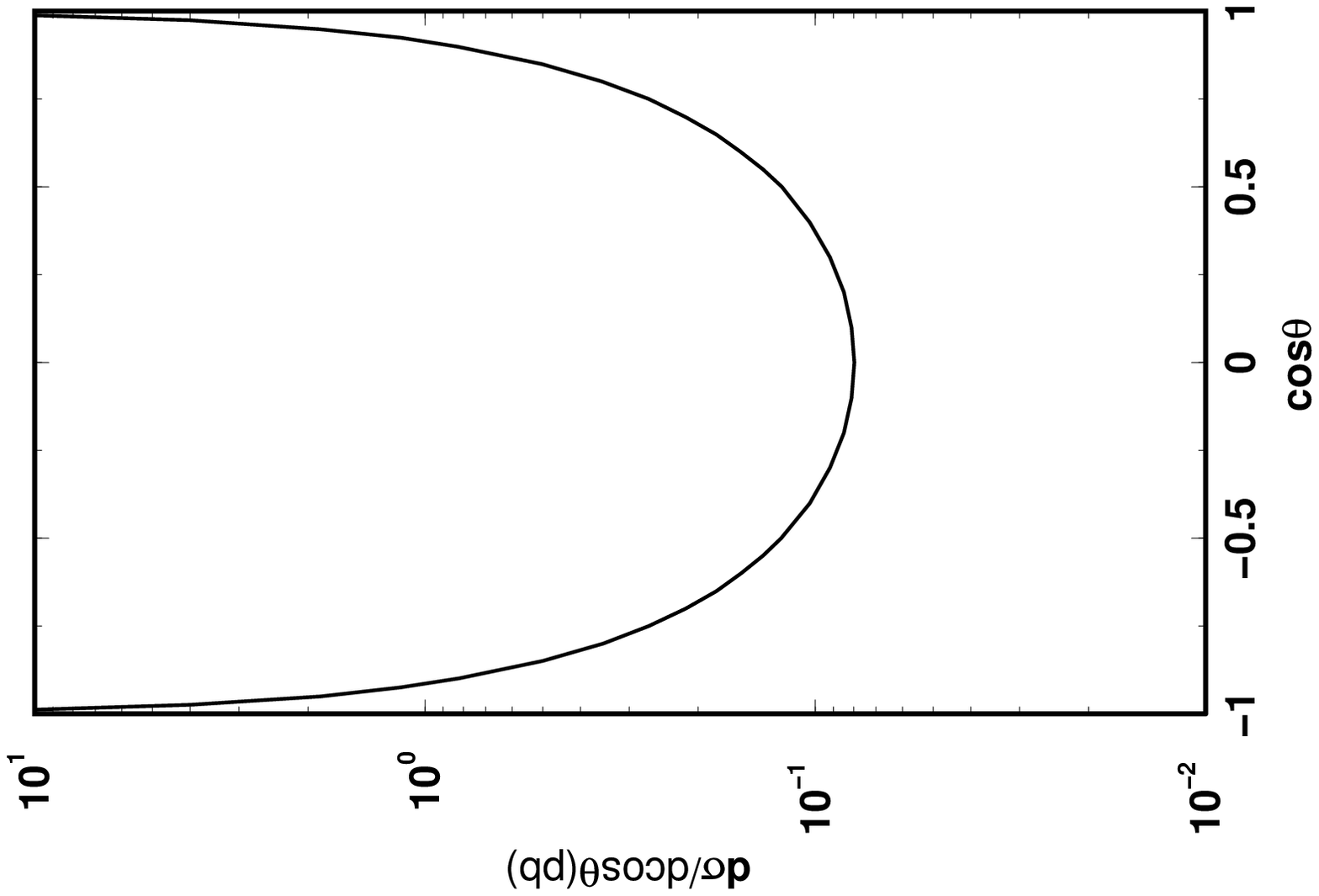,width=4cm,angle=-90} 
\end{tabular}
\caption{The photon distribution
in the laboratory frame at $\sqrt{s}$ = 500 GeV.}
\end{center}
\end{figure}
\noindent
minimum cut-off energy for the photon
to be $5$ GeV.
One can see a rather clear signal from RAZ in the cross section when
the initial proton's helicities are parallel each other.
This is because the helicity distributions of quarks
depend on the spin of the parent protons.
Note that if the proton contains the equal
parton densities of both helicity states as for the unpolarized case,
the convolution completely smears out the RAZ. 

We define also an asymmetry by,
\[  A = \frac{ d \sigma_{-+}/d \cos \theta_{\gamma}
         - d \sigma_{+-}/d \cos \theta_{\gamma}}
       { d \sigma_{-+}/d \cos \theta_{\gamma}
         + d \sigma_{+-}/d \cos \theta_{\gamma}}
       \ ,\]
and plot it in Fig.6.
This asymmetry amounts to $\sim 40$ \%.
\begin{figure}[H]
\begin{center}
\leavevmode\psfig{file=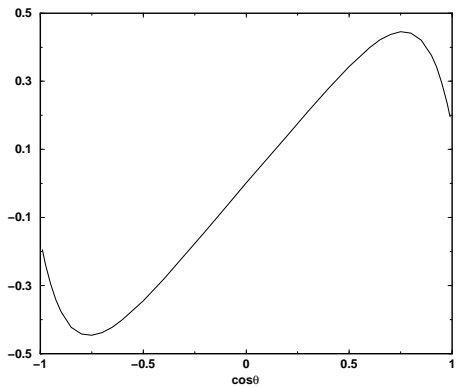,width=6cm,angle=-90}
\caption{Asymmetry with respect to $\cos \theta_{\gamma}$.}
\end{center}
\end{figure}

\section{Discussion}

The RAZ is a prominent feature of the electroweak standard model.
However, it occurs only at the partonic tree level.
Therefore in the experimental measurements,
we must seriously consider the two effects which
possibly wash out the RAZ:
(1) the effects from the PDF (2) radiative corrections.
In this section, we discuss why the experiment at RHIC energy
with polarized beams is appropriate for studying the radiative
weak boson productions.
 
Let us first consider the effects coming from the convolution of the partonic
cross section with the PDFs.
Due to the $V-A$ structure of the $W$ boson interaction,
the partonic process strongly depends on the helicity
states of intial quarks.
However if the proton contains the equal
parton densities of both helicity states as for the unpolarized case,
the convolution smears out the RAZ.
In the case of the polarized beams,
it is known that the helicity distribution of partons strongly
depend on the spin of parent hadrons.
Therefore, the RAZ will not be completely smeared out in the
polarized collisions as shown in Fig.5.
Actually, this was the main idea of ref.~\cite{WSCY}.   
How about is the energy dependence of this smearing effect?
It is easily understood that the dip will be more smeared as
the energy becomes higher.
It is because the small $x$ partons start
to participate in the process
at higher energy and those sea partons carry less
information of the parent proton's spin.
Namely the contribution from the small $x$ partons is expected
to be the same both for the polarized and unpolarized protons.
To verify this feature, we calculate the cross section
at $\sqrt{s}=2000 \ {\rm GeV}$ and plot it in
\begin{figure}[H]
\begin{center}
\begin{tabular}{cc}
\qquad $d \sigma_{+-}/d \cos \theta_{\gamma}$ & 
\qquad $d \sigma_{-+}/d \cos \theta_{\gamma}$ \\
\leavevmode\psfig{file=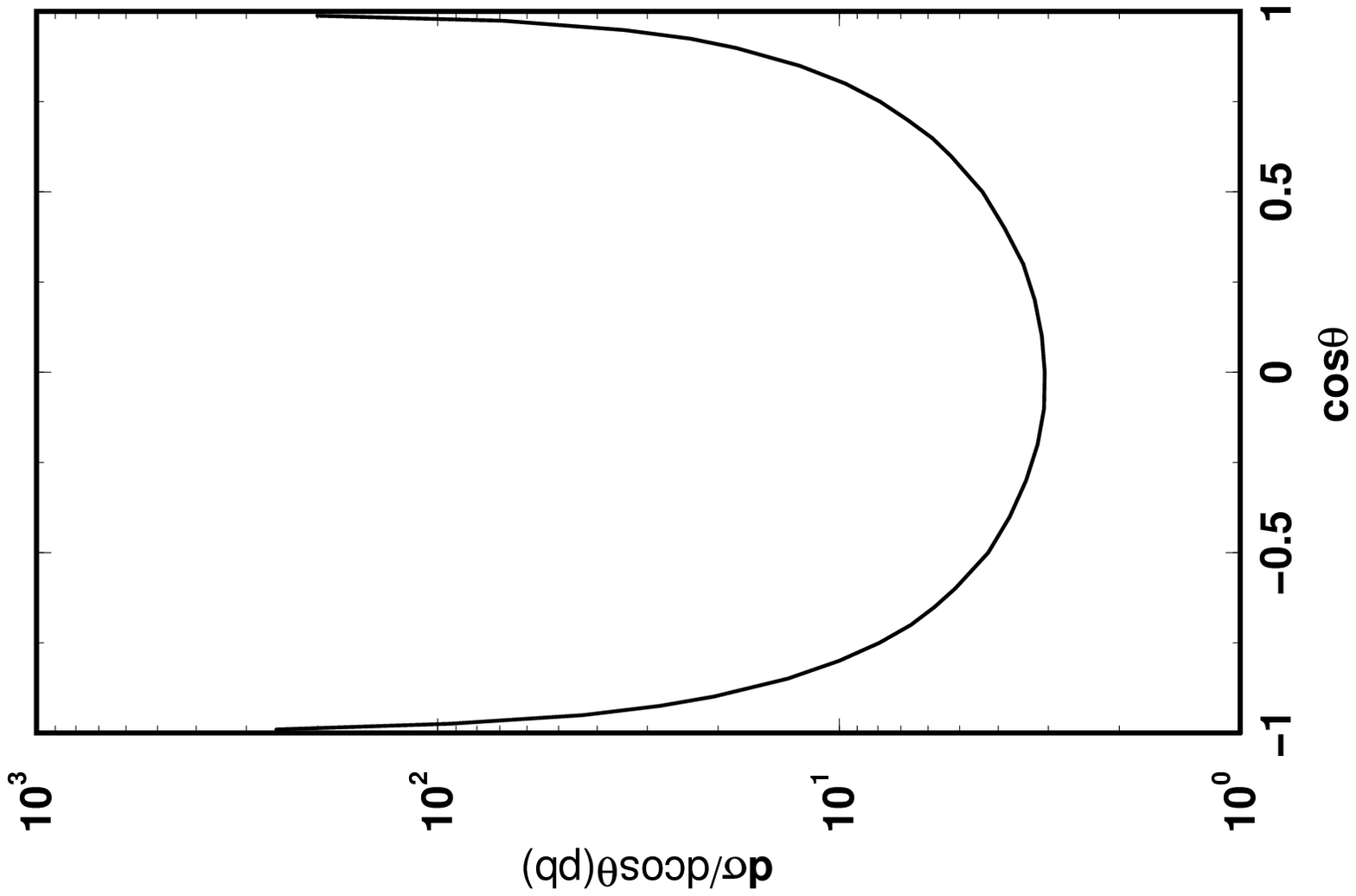,width=4cm,angle=-90} &
\leavevmode\psfig{file=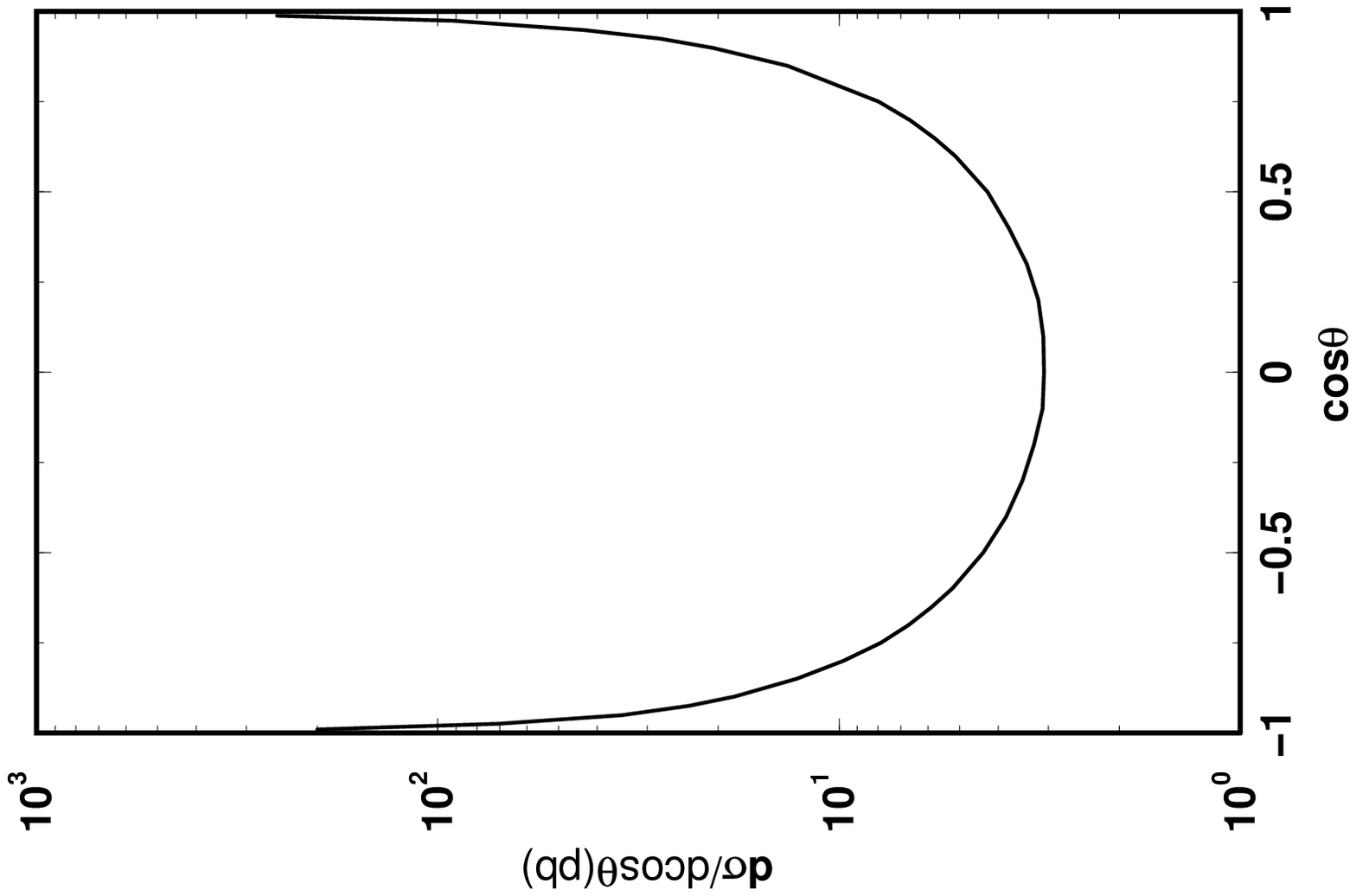,width=4cm,angle=-90}
\end{tabular}
\begin{tabular}{cc}
\qquad $d \sigma_{--}/d \cos \theta_{\gamma}$ &
\qquad $d \sigma_{++}/d \cos \theta_{\gamma}$ \\
\leavevmode\psfig{file=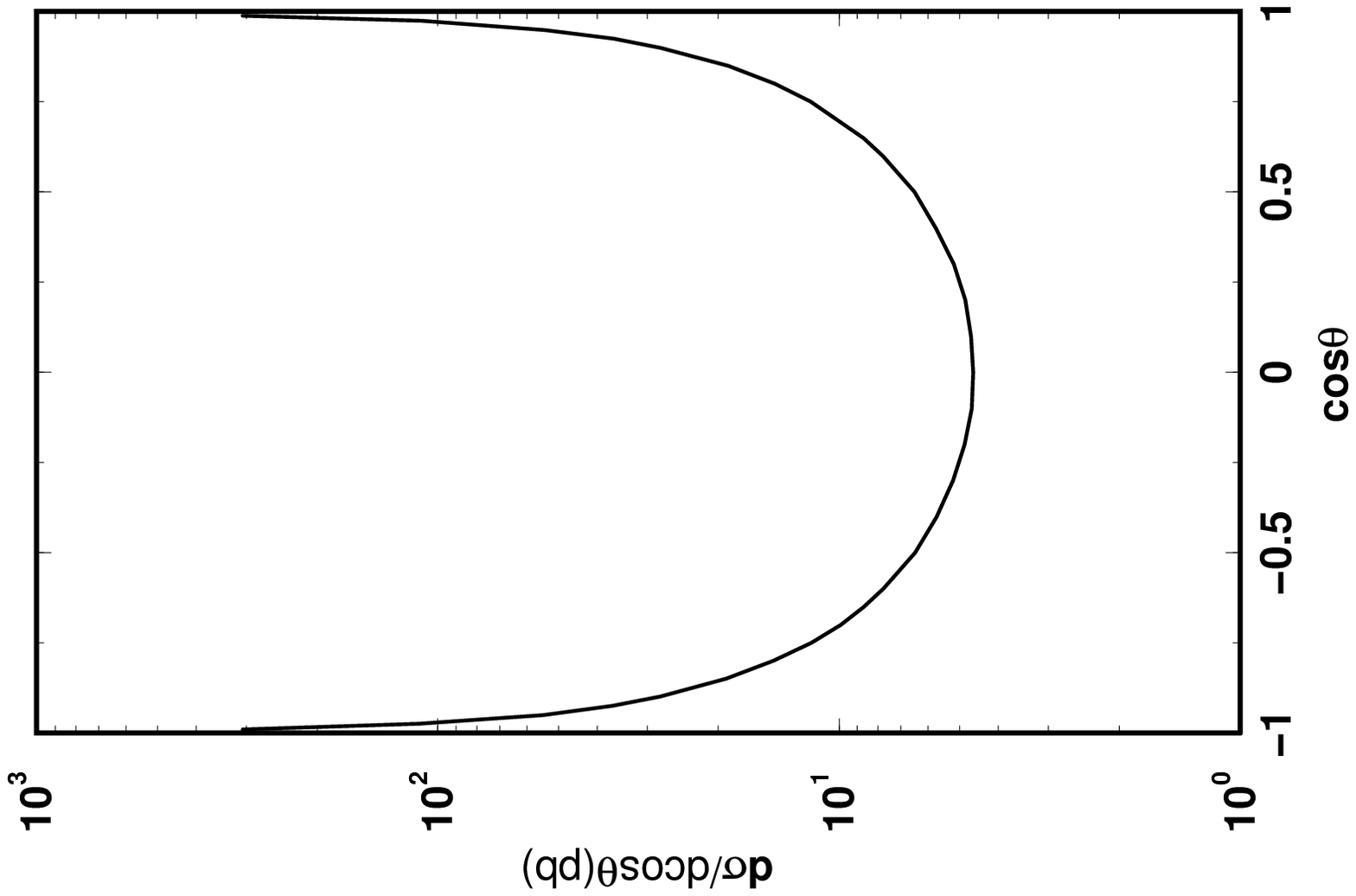,width=4cm,angle=-90} &
\leavevmode\psfig{file=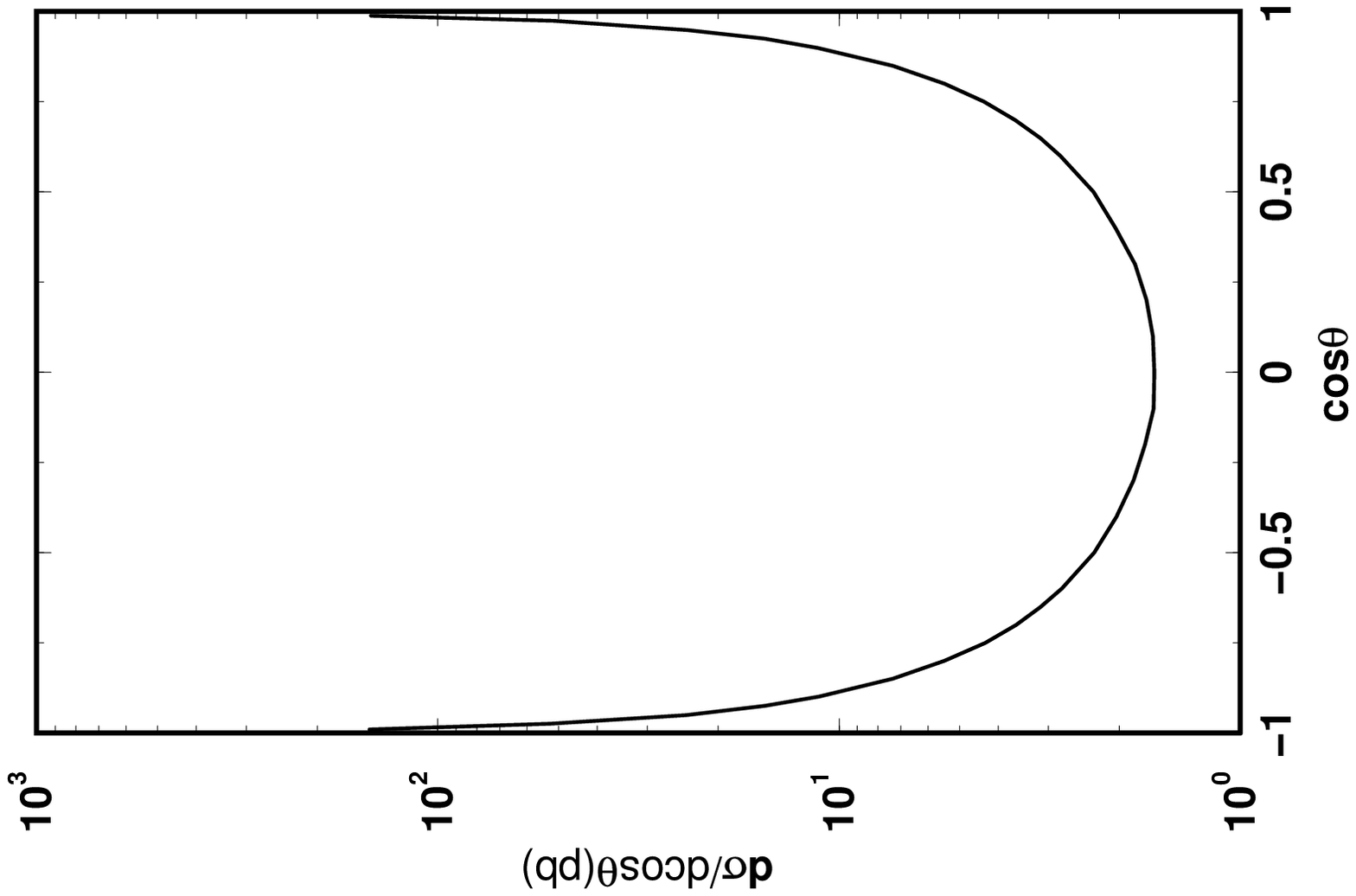,width=4cm,angle=-90} 
\end{tabular}
\caption{The photon distribution in the laboratory frame at $\sqrt{s}$ = 2000 GeV.}
\label{fig7} 
\end{center}
\end{figure}
\noindent
Fig.7.
One can see that the angular distribution become symmetric
for all proton's spin configurations and the RAZ is completely
smeared out.
  
Secondly we discuss the effects of radiative QCD corrections.
In general, since the tree level cross section develops zero,
the radiative corrections are very important.
The QCD corrections to these process were calculated in refs.\cite{SN,O}
and detailed numerical analyses have been done.
The radiative corrections can be classified into three effects:
(1) virtual corrections and soft gluon emissions (2)
hard gluon emissions (3) intial gluon process $q_1 g \to V \gamma q_2$
and $g \bar{q}_2 \to V \gamma \bar{q}_1$.
Among these effects, the first one leaves the RAZ intact.
Its main effect is a constant $K$ factor.
The second and third effects, on the other hand, will completely wash
out the RAZ phenomena.
These effects, however, have been known to be important
at large energies~\cite{SN}.
It is, therefore, expected that in the RHIC energy region, 
it is sufficient to include only the first effect.
As stated above, this effect will not change the shape of the
cross section from the tree level one. 
Therefore the results of the previous section remain the same
qualitatively except for some multiplicative enhancement.
In particular, the asymmetry, Fig.6, remains the same.

\section{Summary} 

We have studied the radiative weak boson production at RHIC energy.
We have pointed out that the experiments in the RHIC energy region
will be very efficient to study this process.
Although the analyses in this article were based on the tree level
calculations, we argued that the higher order QCD corrections
will not smear out the dips from RAZ at intermediate energies. 
Furthermore the polarized beams at RHIC develop these dips.
In the case of $p\, p$ collision, the unpolarized
experiment leads to only the symmetric angular distributions
for the photon.
We can hope that the phenomena, RAZ will survives many
possible smearing effects at the RHIC polarized experiments.

\section*{Acknowledgment}

The work of J.K. was supported in part by the Monbu-kagaku-sho Grant-in-Aid
for Scientific Research No. C-13640289.
Y.K. was supported by the Japan Society for the Promotion of Science.


\end{document}